\documentstyle[12pt,epsf]{article}

\newcommand{\bea}{\begin{eqnarray}}
\newcommand{\eea}{\end{eqnarray}}
\newcommand{\ba}{\begin{array}}
\newcommand{\ea}{\end{array}}
\newcommand{\be}{\begin{equation}}
\newcommand{\ee}{\end{equation}}
\newcommand{\beas}{\begin{eqnarray*}}
\newcommand{\eeas}{\end{eqnarray*}}
\def\identity{{\rlap{1} \hskip 1.6pt \hbox{1}}}

\newcommand{\nbox}{{\,\lower0.9pt\vbox{\hrule \hbox{\vrule height 0.2 
cm \hskip
0.2 cm \vrule height 0.2 cm}\hrule}\,}}

\DeclareFixedFont{\xiiss}{OT1}{cmss}{m}{n}{12}
\DeclareFixedFont{\ixss}{OT1}{cmss}{m}{n}{9}
\DeclareFixedFont{\cmrnine}{OT1}{cmr}{m}{n}{9}

\newcommand{\CC}{\hbox{\xiiss C\kern-.4emI}}
\newcommand{\RR}{\hbox{\xiiss R\kern-.45emI}}
\newcommand{\ZZ}{\hbox{\xiiss Z\kern-.4emZ}}
\newcommand{\CCs}{\hbox{\ixss C\kern-.4emI}}
\newcommand{\ZZs}{\hbox{\ixss Z\kern-.4emZ}}
\newcommand{\pa}{\partial}

\newcommand{\tr}{{\rm tr}\ }
\newcommand{\pasl}{\pa\kern-.55em /}
\def\href#1#2{#2}

\begin{document}

\begin{titlepage}

\begin{flushright}
SU-ITP 01-10\\
SLAC-PUB-8796\\
hep-th/0103179
\end{flushright}

\vskip 2.5cm

\begin{center} {\Large \bf

Quantum Hall Physics \\

Equals\\
 Noncommutative Field Theory

                                                         } 
\end{center}

\vspace{1ex}

\begin{center}
{\large Simeon Hellerman$^{a,b}$ and Mark Van Raamsdonk$^b$}

\vspace{5mm}
{${}^a$ \sl Stanford Linear Accelerator Center} \\
{\sl Stanford, CA 94305 U.S.A.} \\
{\tt simeon@itp.stanford.edu} 

\vspace{5mm}
{${}^b$ \sl Department of Physics} \\
{\sl Stanford University} \\
{\sl Stanford, CA 94305 U.S.A.} \\
{\tt mav@itp.stanford.edu}

\end{center}

\vspace{2.5ex}
\medskip
\centerline {\bf Abstract}

\bigskip

In this note, we study a matrix-regularized version of non-commutative 
$U(1)$  Chern-Simons theory proposed recently by Polychronakos. We 
determine a complete minimal basis of exact wavefunctions for the 
theory at arbitrary level $k$ and rank $N$ and show that these are in 
one-to-one correspondence with Laughlin-type wavefunctions describing
excitations of a quantum Hall droplet composed of $N$ electrons at filling fraction 
$1/k$. The finite matrix
Chern-Simons theory is shown to be precisely equivalent to the theory
of composite
fermions in the lowest Landau level, believed to provide an accurate
description of the filling fraction $1/k$ fractional quantum Hall state. 
In the large $N$ limit, this implies that level $k$ 
noncommutative $U(1)$ Chern-Simons theory is equivalent to the
Laughlin theory of the filling fraction $1/k$ quantum Hall fluid, as 
conjectured recently by Susskind.

\bigskip

\end{titlepage}

\newpage

\section{Introduction}

In its simplest form, noncommutative geometry is characterized by a
pair of coordinates which do not commute in the same sense that
canonically conjugate positions and momenta do not commute in quantum
mechanics, 
\be
\label{comm}
[x,y] = i \theta \; ,
\ee
where $\theta$ is a dimensionful parameter. Over the past few years,
it has become increasingly apparent that noncommutative geometry plays
an important role in string theory \cite{cds,sw}. However, such a
commutation relation also appears in a much simpler context, that of
charged particles in a strong magnetic field. 

For a single particle of unit charge moving in two dimensions under the 
influence of a constant magnetic field, the Lagrangian is given by
\be
\label{lag}
L = {m \over 2} (\dot{x}^2 + \dot{y}^2) + {1 \over 2} B 
(\dot{x} y - \dot{y} x) .
\ee
It is well known that the eigenstates of this system lie in Landau 
levels, degenerate sets of states at energies given by 
$E_n = (n + {1 \over 2}){ B \over m}$. In each level, the degeneracy 
is equal to one state for each unit of area defined by the inverse 
density of flux quanta. In the limit of very large magnetic field,
particles are restricted to the lowest Landau level, and the 
physics may
be described by ignoring the first term in (\ref{lag}). Canonically
quantizing the resulting system, we are led to precisely the
commutation relation (\ref{comm}) with $\theta$ replaced by ${1 \over
B}$. Thus, the two dimensional coordinate space becomes the phase
space for the system and gives a simple realization of a
noncommutative geometry.

The physics of electrons in the lowest Landau level exhibits many 
fascinating properties. In particular, when the electron density lies 
at certain rational fractions of the density corresponding to a fully 
filled lowest Landau level, the electrons condense into special 
incompressible fluid-like states whose excitations exhibit such unusual 
phenomena as fractional charge and fractional statistics. These states 
also have a gap in their excitation spectrum which gives rise to the 
experimentally observed fractional quantum Hall effect (for a recent 
review, see \cite{girvin} and references therein). For the filling 
fractions $1/k$, the physics of these states is accurately described by 
certain wavefunctions proposed by Laughlin, and more general 
wavefunctions may be written down describing the various types of 
excitations about the Laughlin states.     

Given that the first quantized system of a particle in a strong 
magnetic field naturally realizes a noncommutative space, it is 
interesting to speculate that the second-quantized field theory 
description of the Quantum Hall fluid for various filling fractions 
might 
involve a noncommutative field theory. In a recent paper by Susskind 
\cite{susskind}, 
this possibility was put forward in a precise conjecture, that the 
Laughlin state of electrons at filling fraction $1/k$ is precisely 
described by the noncommutative version of $U(1)$ Chern-Simons theory 
at level $k$. Subsequently, Polychronakos \cite{poly} proposed 
that a particular 
matrix regularized version of this level $k$ noncommutative 
Chern-Simons theory could be used to describe a finite number of electrons in 
a ``droplet'' of the Quantum Hall fluid at filling fraction $1/k$.  

In this paper, we cement the connection between noncommutative 
Chern-Simons theory and the Laughlin theory of the Quantum Hall fluid. 
Specifically, write down the complete set of exact wavefunctions of the 
Polychronakos model and show that they are in one-to-one correspondence 
with wavefunctions describing excitations of a filling fraction $1/k$ 
Quantum Hall droplet in the Laughlin theory (the Hilbert spaces are 
isomorphic and the energy levels are the same). By taking the limit of 
a large number of electrons, this should imply that the filling 
fraction $1/k$ Quantum Hall fluid of infinite extent is well described by 
the noncommutative $U(1)$ Chern-Simons theory at level $k$, as 
conjectured by Susskind. 

The plan of the paper is as follows. In section 2, we review some basic 
properties of charged fermions in the lowest Landau level. We review 
Laughlin's wavefunctions for the filling fraction $\nu = 1/k$ states as 
well as the more general wavefunctions describing excitations about 
these states. In section 3, we describe the noncommutative Chern-Simons 
theory and its matrix description, and review some of the evidence put 
forward by Susskind that this theory describes the fractional quantum 
Hall fluid. In section 4, we present the finite dimensional matrix 
Chern-Simons theory proposed by Polychronakos and review its 
quantization. Section 5 contains the main results of this paper: we 
write down explicitly the wavefunctions for a complete minimal basis of 
energy eigenstates for the theory and show that they are in one-to-one 
correspondence with wavefunctions describing excitations of a quantum 
Hall droplet described by the Laughlin theory at $1/k$. In section 6, 
we offer some concluding remarks. 

\section{Charged particles in the lowest Landau level}

In this section, we review some aspects of the physics of charged 
fermions in a strong magnetic field. We recall Laughlin's wavefunctions 
describing particles in the lowest Landau level at filling fractions 
$1/k$ as well as the more general wavefunctions describing excitations 
about these states. 

We begin by considering a single particle of mass $m$ and unit
charge in a magnetic field $B$. It is convenient to break the
degeneracy arising from translation invariance, so we also include a 
harmonic oscillator potential (which we may take to be arbitrarily 
weak). The complete Lagrangian is then given (in the radial gauge $A_i 
= {1 \over 2} B \epsilon^{ij} x^j)$ by
\be
\label{lg1}
L = {m \over 2} \dot{x}^2_i + {B \over 2} \epsilon^{ij} \dot{x}_i x_j
- {1 \over 2} \kappa x_i^2 \; .
\ee
The energy eigenstates for this system may be determined exactly. In
the limit $m \to 0$ (equivalent to strong magnetic field), the 
wavefunctions whose energies remain finite relative to the ground state 
energy may be
labeled by an integer $n \ge 0$ and are given by
\be
\label{more}
\langle\vec{x}|n\rangle = \sqrt{B^{n+1} \over 2^{n+1} \pi n!} z^n e^{-
{B \over
4}|z|^2}
\ee
where $z = x +iy$. These have energies $E_n = {\kappa \over B} (n + {1 
\over 2} )$. 

Another way to arrive at these states is
by taking $m=0$ at the start. Then the canonical commutation relations
give
\be
\label{com2}
[x,y] = {i \over B}
\ee
and the Hamiltonian is that of a Harmonic oscillator,
\[
H =  {\kappa \over B} (a^\dagger a + {1 \over 2}) 
\]
where we have defined $a = \sqrt{B \over 2}(x + i y)$ so that $a$ and
$a^\dagger$ have the usual commutation relations of creation and
annihilation operators. The eigenstates, again labeled by $n$, are of
course
\be
\label{onew}
|n \rangle 
 = {1 \over \sqrt{n!}} (a^\dagger)^n |0 \rangle 
\ee
with energies as above. 

It will be helpful to understand the precise relationship between these 
wavefunctions (\ref{onew}), which depend only on a single variable, and 
those in (\ref{more}) which depend on two coordinates \cite{djt}. In the reduced
system, we 
may define a coherent state basis $|z \rangle $ by $a|z \rangle  = z|z 
\rangle $. In this 
basis, we have $\langle z|n \rangle  = {1 \over \sqrt{n!}} z^n$, so we
see that the 
analytic part of the lowest Landau level wavefunctions (\ref{more}) 
should be identified with the coherent state wavefunction 
$\langle z|\psi \rangle $ in 
the system with reduced phase space.  

In the second picture, it is valuable to note that the hamiltonian $H$ 
is proportional to the radius squared operator, $H = {1 \over 2} \kappa 
R^2$ and also generates rotations (because of the commutation relation 
(\ref{com2})). Thus, the state labeled by $n$ has energy, angular 
momentum and $R^2$ all proportional to $n$.
  
Consider now the system of $N$ fermions, each described by the
Lagrangian (\ref{lg1}) in the limit of large magnetic field (we assume
for now that their interactions may be ignored). The wavefunctions must 
be completely antisymmetric and it is easy to see that an orthogonal 
basis of energy eigenstates is provided by the set of wavefunctions
\be
\label{wone2}
|\{n_i\} \rangle  = \epsilon^{i_1 \cdots i_N} a^\dagger_{i_1}{}^{n_1} \cdots
a_{i_N}^\dagger{}^{n_N}|0 \rangle  \; ,
\ee
where $n_1 < n_2 < \cdots < n_N$. These correspond to states with one 
particle in each of the energy levels labeled by $n_i$, so the total 
energy is $E = {\kappa \over B} ({N \over 2} + \sum n_i)$. In the 
original two dimensional language, these may be written as
\be
\label{wone}
\langle \vec{x}|\{n_i\} \rangle  = \epsilon^{i_1 \cdots i_N} z_{i_1}^{n_1} 
\cdots
z_{i_N}^{n_N} e^{-{B \over 4}\sum|z_i|^2} \; .
\ee 

The ground state corresponds to choosing $n_i = i - 1$, so that the 
$R^2$ expectation value for the outermost particle is $R^2 = {2  
\over B} (N - {1 \over 2})$. The disc bounded by this radius contains 
$\sim N$ quanta of flux (the density of flux quanta is $B/{2 \pi}$), so 
we see that the ground state corresponds to a circular ``quantum hall 
droplet'' with filling fraction $\nu = 1$. 
Since the number of states per unit area in the lowest Landau level is 
equal to the number of flux lines per unit area, this ground state 
droplet is maximally dense and therefore incompressible. This may be 
seen directly by noting that the size of the droplet is independent of 
the strength of the harmonic oscillator potential. As a result, the 
external potential we have introduced should not significantly affect 
the physics of the droplet. 

Using Fact 1 from the appendix, we may rewrite the $\nu = 1$ ground 
state wavefunction $|\psi_1 \rangle  \equiv |{n_i = i-1} \rangle $ in a more 
standard form,
\be
\label{normal}
\langle \vec{x}| \psi_1 \rangle  = \left( \prod_{i<j}^N (z_i - z_j) \right) 
e^{-{B \over 4}\sum|z_i|^2} \; . 
\ee
As required by antisymmetry, the wavefunction vanishes as any two 
particles become coincident. This also ensures that the wavefunction 
will continue to provide a good description in the presence of 
repulsive interparticle interactions as long as they are sufficiently 
weak/short-ranged. 

Systems of electrons in a strong magnetic field also condense into 
special incompressible states at certain rational filling fractions, 
giving rise to the fractional quantum Hall effect. Generalizing the 
$\nu = 1$ ground state wavefunction (\ref{normal}), Laughlin \cite{laughlin}
proposed that for $\nu = 1/k$, these fractional Quantum hall states could be 
accurately described by wavefunctions 
\bea
\langle \vec{x}|\psi_{1/k}\rangle &=& \left(\prod_{i<j}^N (z_i - 
z_j)^k 
\right) e^{-{B \over
4}\sum|z_i|^2}\nonumber\\
&=& \left( \epsilon^{i_1 \cdots i_N} z_{i_1}^0 \cdots
z_{i_N}^{N-1} \right)^k  e^{-{B \over 4}\sum|z_i|^2} \; .
\label{frac}
\eea
It is easy to see that the highest power of a given $z_i$ appearing 
(proportional to the radius squared of the droplet) is $k(N-1)$, so 
this state has an electron density of $1/k$ times the $\nu = 1$ ground 
state, as required. The presence of an additional factor $(z_i - 
z_j)^{k-1}$  for each pair of particles relative to the $\nu=1$ state 
(and therefore an additional phase of $(-1)^{k-1}$ when the particles 
are interchanged) may be understood to arise from interparticle 
interactions which result in the binding of $k-1$ flux lines to each 
electron, forming a ``composite fermion'' \cite{jain,jainbook}. This helps to 
minimize the 
repulsive Coulomb potential, as is evident from the rapid vanishing of 
the wavefunction as any two particles approach each other. The 
resulting composite fermions are weakly interacting and feel a reduced 
net magnetic field corresponding to the unattached flux lines. The 
density of composite fermions is precisely equal to the density of 
these unattached flux lines, so the $\nu = 1/k$ state may be 
interpreted as the $\nu=1$ state of the composite fermions. It is 
important to note that $k$ must be an odd integer in order to preserve 
the fermionic statistics of the particles (reflected in the 
antisymmetry of the wavefunction).

Based on this composite fermion picture, the general set of 
wavefunctions describing excitations of the $\nu = 1/k$ state should be 
obtained from the $\nu=1$ wavefunctions (\ref{wone}) simply by 
multiplying by the factor $\prod (z_i - z_j)^{k-1}$ corresponding to 
the flux attachment. Equivalently, the wavefunctions describing 
excitations of the $1/k$ state may be taken to be the subset of energy
 eigenstates states spanned by the basis (\ref{wone}) which contain a zero of
 order at least $k$ when any pair of particles become coincident.  

Using the harmonic oscillator representation (which will be most 
convenient for our later comparison), a minimal basis of wavefunctions 
describing excitations of the $\nu = 1/k$ state in the Laughlin theory 
is therefore given by
\be
\label{ba1}
|\{n_i\}, k \rangle = \epsilon^{i_1 \cdots i_N} a_{i_1}^\dagger{}^{n_1} 
\cdots
a_{i_N}^\dagger{}^{n_N} \left( \epsilon^{i_1 \cdots i_N} 
a^\dagger_{i_1}{}^0 \cdots a_{i_N}^\dagger{}^{N-1} \right)^{k-
1}|0\rangle
\ee
where $n_i$ are integers such that $0 \ge n_1 < \cdots < n_N$. The 
energies of these states is given by $E(\{n_i\},k) = {\kappa \over B} 
({N \over 2} (1 + (k-1)(N-1)) + \sum n_i)$ 

Using Fact 2 from the appendix, it is not difficult to see that another 
minimal basis of wavefunctions is given by
\be
\label{ba2}
|\{c_i\}, k \rangle = (\sum a^\dagger_i{}^N)^{c_N} \cdots (\sum 
a^\dagger_i 
{})^{c_1} \left( \epsilon^{i_1 \cdots i_N} a_{i_1}^\dagger{}^0 \cdots 
a_{i_N}^\dagger{}^{N-1} \right)^{k}|0\rangle
\ee
where $c_i$ are arbitrary non-negative integers and the energy of these 
states is $E = {\kappa \over B} ({N \over 2}(1 + k(N-1))  + \sum m 
c_m)$.

In section 5, we will construct two bases of the wavefunctions for the 
finite $N$ matrix noncommutative Chern-Simons theory and see that they 
have a structure almost identical to the two bases constructed here. In 
preparation for this, we provide a brief review in the next two 
sections of the noncommutative Chern-Simons theory, its matrix 
decription, and the finite dimensional matrix version proposed by 
Polychronakos.

\section{Noncommutative Chern-Simons theory}

Chern-Simons theory with gauge group U(1) on noncommutative
two-dimensional space is described by an action 
\[
S = {k \over 4 \pi} \int d^3 x \epsilon^{\mu \nu \lambda}
\left(A_\mu \star \partial_\nu A_\lambda + {2 \over 3} A_\mu \star 
A_\nu \star A_\lambda \right) \; .
\]
where 
\[
f \star g  = e^{{i \over 2} \theta \epsilon^{ij} \partial^f_i 
\partial^g_j}fg 
\]
This theory is invariant under arbitrary  noncommutative gauge
transformations (trivial at infinity) given by
\[
A_{\mu} \to U^{-1} \star A_{\mu} \star U + i U^{-1} \star
\partial_\mu U 
\]
as long as $k$ (the level) is an integer \cite{np,blp,sjlev}. Note that the 
appearance of a cubic term in the action and the quantization of the 
level occurs even for the case of $U(1)$, unlike in the commutative 
theory. 

The theory may be written in an equivalent form by choosing the gauge
$A_0 = 0$ in which case the action becomes
\be
\label{lag2}
S = {k \over 4 \pi} \int d^3 x \epsilon^{ij} A_i \partial_t A_j 
\ee
while the equation of motion for $A_0$ must be imposed as a constraint, 
\be
\label{con}
F_{ij} = \partial_i A_j - \partial_j A_i -i A_i \star A_j + i A_j
\star A_i = 0
\ee

It turns out that this action and constraint also arise from a matrix
model in $0+1$ dimensions, given by 
\be
\label{mat}
S = {k \over \theta} \int dt {\rm Tr} \left({1 \over 2}
\epsilon^{ij} D X^i X^j \right)  + k \int dt {\rm Tr}(A) 
\ee
where X and A are hermitian matrices \cite{bbst,poly2,kluson}. The 
action is invariant under gauge transformations
\[
X^i \to U^{-1} X^i U, \qquad \qquad A \to U^{-1} A U +i U^{-1}
\partial_t U 
\]
as long as $U$ is taken to be trivial at $t=\pm \infty$ and $k$ is an
integer. We may choose the gauge $A=0$ in which case the action
becomes 
\be
\label{max}
S = {k \over \theta} \int dt {\rm Tr} \left({1 \over 2}
\epsilon^{ij} \dot{X}^i X^j \right) 
\ee
while the equation of motion for A is 
\be
\label{mcon}
[X^1,X^2] = i \theta \identity
\ee
which must be taken as a constraint (note that the commutator here is a 
matrix commutator). This has no solutions for finite dimensional 
matrices (as may be seen by taking the trace), thus, we must take $X^i$ 
to be infinite dimensional. A particular solution is $X^i = y^i$, where 
$y^1$ and $y^2/ \theta$ are the usual matrices representing $x$ and $p$ 
in the harmonic oscillator basis. Expanding the action (\ref{max}) and 
constraint (\ref{mcon}) about this classical solution, $X^i = y^i + 
\theta \epsilon^{ij} A_j$ and considering $A_j$ to be functions of the 
noncommuting coordinates $y^i$ gives precisely the Lagrangian 
(\ref{lag2}) with the constraint (\ref{con}).\footnote{This step 
requires making the transition from the operator formalism for fields 
on noncommutative space to the representation in terms of ordinary 
functions multiplied by the star products. Explicitly, we have
\[
[y_i, f] \to i \theta \epsilon_{ik} \partial_k f \; , \qquad {\rm 
Tr}(f_1 \cdots f_n) \to {1 \over 2 \pi \theta} \int d^2 x (f_1 \star 
\cdots \star f_n)
\]
For more details of the relationship between these two representations, 
see for example \cite{harvey}.} Thus, the matrix model (\ref{mat}), 
expanded about the background $X^i = y^i$ is equivalent to the 
noncommutative $U(1)$ Chern-Simons theory. To be precise, we should 
restrict to fluctuations $A_i$ about the background described by 
compact operators (this is equivalent to the condition that the fields 
vanish at infinity in the field theory language) with a similar 
condition on the allowed gauge transformations \cite{harvey}. 
Interestingly, the same matrix model expanded about a different 
background $X^i = y^i \otimes \identity_{N \times N}$ gives the $U(N)$ 
noncommutative Chern-Simons theory (this highlights the need to 
restrict the allowed $X$'s to a specific subset of all hermitian 
matrices before the matrix model gives a well defined theory). 
  
Starting from the matrix action (\ref{max}), canonical quantization 
gives the commutation relations
\[
[X^1_{mn} , X^2_{pq}] =  i { \theta \over k} \delta_{mq} \delta_{np}
\; .
\]
Thus, we may choose the wavefunction to be a function of the matrix $X
= X^1$ and represent $X^2$ as
\[
X_{mn}^2 = {\theta \over k} i {\partial \over \partial X_{nm}} \; .
\]
The constraint (\ref{mcon}) should then be imposed as an operator 
constraint on the wavefunction. With the proper operator ordering, the 
left side of (\ref{mcon}) generates $U(N)$ transformations on the 
wavefunction \cite{susskind}, and it follows that the constraint is 
equivalent to  
the condition  
\be
\label{master}
\Psi(U^{-1} X U) = ({\rm det}U)^k \Psi(X) \; .
\ee
Since the Hamiltonian for the theory is zero, this equation provides
all the information about the allowed wavefunctions.\footnote{However, 
as discussed above, we must be careful to restrict to an appropriate 
subset of allowed $X$'s and $U$'s before the theory is well defined.}  
 
Because the matrices are infinite dimensional, it is difficult to 
determine solutions to this constraint or even to understand whether it 
has well defined solutions without some sort of regularization. 
However, as pointed out in \cite{susskind} this constraint already 
provides some direct evidence for a connection with the Laughlin 
states. Specifically, if $U$ is taken to be a permutation matrix which 
would permute two elements of a diagonal matrix, then we have ${\rm 
det}(U) = -1$, so (\ref{master}) indicates that the wavefunction obeys 
fermionic statistics for odd values of $k$ and bosonic statistics for 
even values of $k$. Identifying $k$ with the inverse filling fraction 
in the Laughlin theory, this gives precisely the relation between 
statistics and filling fraction obeyed in the Laughlin wavefunctions.

In order to proceed further, it is very helpful to work with a 
regulated version of the theory that is consistent with finite 
dimensional matrices. Such a theory has been proposed by Polychronakos 
\cite{poly}. In the next section, we review his construction, and in 
section 5 determine the complete set of allowed wavefunctions for the 
theory.

\section{Finite N Matrix Chern-Simons theory}

In \cite{poly}, Polychronakos proposed a modification of the action 
(\ref{max}) to make it consistent with matrices of arbitrary finite 
dimension $N$. He suggested that the resulting action should provide a 
description of the states of a quantum hall droplet of finite extent 
composed of $N$ electrons. Further, he showed that many of the expected 
excitations of a quantum hall droplet, including area-preserving 
boundary excitations and quasiholes with fractional charge $1/k$, 
appear naturally in the model.

The action proposed by Polychronakos is 
\[
S = {k \over 2\theta} \int dt {\rm Tr} \left( \epsilon^{ij} D X^i X^j - 
\omega X_i^2 \right) + i \int dt \Psi^{\dagger} D \Psi + k \int dt {\rm 
Tr}(A) \; .
\]
Relative to the orginal matrix model (\ref{max}), there are two 
additional terms. The first is a potential term $-\omega 
X_i^2$ analogous to the harmonic oscillator potential considered in 
section 2. Again, this serves to break the degeneracy arising from 
translation invariance and also provides a Hamiltonian for the theory 
that selects a unique ground state. 

The other new term involves complex bosons $\Psi_i$ transforming in the 
fundamental representation of the gauge group. With the additional
$\Psi_i$  term, the 
constraint becomes
\be
\label{psicon}
[X^1, X^2]_{mn} + {i \theta \over k} \Psi_m \Psi^{\dagger}_n = i
\theta \delta_{mn} \; .
\ee
This is now consistent with finite dimensional $X^i$ since the left 
side is no longer traceless. For $N \times N$ matrices the trace of 
this equation gives
\be
\label{trcon}
\Psi^{\dagger}_m \Psi_m = Nk \; .
\ee
As demonstrated in \cite{poly} the constraints may be explicitly solved 
classically by using the $U(N)$ symmetry to make $X^1$ diagonal and 
$\Psi$ real. The resulting system, with no remaining gauge symmetry 
apart from the permutations, contains $N$ real degrees of freedom, the 
same number as $N$ electrons in the lowest Landau level.\footnote{In 
fact, as discussed in \cite{poly} the theory obtained in this way is 
precisely the Calogero model for $N$ particles in one dimension.}  

As discussed in \cite{poly}, a useful way to understand the
relationship of this theory to the original Chern-Simons theory is to
note that in a particular gauge, we
may solve the trace part of constraint (\ref{trcon}) by taking 
\[
\Psi_i = (0,0,\cdots,\sqrt{kN}) \; .
\]
The resulting theory, with the harmonic oscillator potential taken to
zero, retains a $U(N-1) \times U(1)$ symmetry, and is described by the
Lagrangian (\ref{max}) and the constraint (\ref{mcon}) modified so
that the $(N,N)$ entry of the right hand side becomes
$i\theta(1-N)$. This is essentially the minimal modification of the
original matrix Chern-Simons theory to make it consistent with finite 
dimensional matrices. In the $N=\infty$ limit, the constraint equation
formally goes over to the original one (\ref{mcon}) so it seems
reasonable to claim that the $N=\infty$, $\omega \to 0$ limit of the
Polychronakos model is noncommutative Chern-Simons
theory. Alternately, we may think of this limit as a particular way of
giving a well defined definition of non-commutative Chern-Simons theory.
    
Let us now consider the finite matrix Chern-Simons theory quantum
mechanically, following \cite{poly}. Rather than solving the
constraint before quantization,
 we will proceed 
by first quantizing the theory and then applying the constraints as 
operator conditions on the wavefunction. It is convenient to introduce 
the matrix $A \equiv \sqrt{k \over 2 \theta}  (X + i Y) $. Then the 
canonical commutation relations give 
\[
[A_{ij}, A^{\dagger}_{kl}] = \delta_{il} \delta_{jk}  \qquad \qquad 
[\Psi_i, \Psi^{\dagger}_j] = \delta_{ij} \; .
\]
Thus, we have $N^2$ creation and annihilation operators coming from the 
$A$'s and $N$ creation and annihilation operators coming from the 
$\Psi$'s. The hamiltonian is given by
\be
\label{ham}
H = \omega ({N^2 \over 2} + {\cal N}_A)
\ee
where ${\cal N}_A = \sum A^{\dagger}_{ij} A_{ji}$ is the number 
operator associated with the $A$'s. Thus, energy eigenstates will be 
linear combinations of terms with a fixed number of $A^{\dagger}$  
creation operators acting on the Fock space vacuum. Examining the trace 
part of the constraint (\ref{trcon}), we see that the left hand side is 
simply the number operator ${\cal N}_\Psi$ associated with the $\Psi$ 
creation and annihilation operators, so all wavefunctions must have a 
fixed number $Nk$ of $\Psi^{\dagger}$ creation operators acting on the 
Fock space vacuum. Finally, as shown in \cite{poly}, the traceless part 
of the constraint (\ref{psicon}) demands that the wavefunction be 
invariant under $SU(N)$ transformations, under which the creation 
operators transform in the adjoint and antifundamental,  
\be
\label{gauge}
A^{\dagger} \to U A^{\dagger} U^{\dagger} \; , \qquad \qquad 
\Psi^{\dagger} \to \Psi^{\dagger} U^{\dagger} \; .
\ee
Therefore, as pointed out in \cite{poly} the problem reduces to a group 
theory problem of determining all ways of combining $Nk$ 
antifundamentals (symmetrized) and any fixed number of adjoints to form 
a singlet of $SU(N)$. In the next section we solve this problem 
explicitly and write down the complete set of wavefunctions for the 
model. We will see that they bear a striking resemblance to the 
Laughlin-type wavefunctions considered in section 2 and then argue that 
the two systems are in fact equivalent.

\section{Determining the wavefunctions}

From the results of the previous section, any wavefunction describing 
an energy eigenstate will be a sum of terms of the form 
\[
A^{\dagger}{}^{i_1}_{j_1} \cdots A^{\dagger}{}^{i_M}_{j_M} 
\Psi^{\dagger}{}_{l_1} \cdots \Psi^{\dagger}{}_{l_{Nk}} |0\rangle
\]
with $Nk$ creation operators $\Psi^\dagger$ and a fixed number $M$ of 
$A^\dagger$ creation operators acting on the Fock space vacuum, such 
that the set of terms forms a singlet under the $SU(N)$ transformations 
(\ref{gauge}). Here, we have written fundamental indices as upper 
indices and antifundamental indices as lower indices. The problem of 
forming a singlet out of the $Nk$ antifundamentals and $M$ adjoints is 
basically just a matter of contracting all the indices using the 
invariant tensors of $SU(N)$. In particular, we may contract any upper 
index with any lower index using $\delta^i_j$, or we may contract any 
set of N lower indices (or N upper indices) with an $N$-index 
completely antisymmetric $\epsilon$ tensor. We may place the further 
restriction that only one type of epsilon tensor may appear (either 
upper or lower index) since the product of an upper index epsilon 
tensor and a lower index epsilon tensor may be rewritten as a sum of 
products of $2N$ delta functions. 

Let us consider first the indices on the $Nk$ $\Psi^\dagger$s. The 
lower index on each $\Psi^\dagger$ must contract either with the upper 
index on an $A^\dagger$ or with an epsilon tensor. If the 
$\Psi^\dagger$ contracts with an $A^\dagger$, the resulting object will 
again have a single lower index. Repeating this logic, we conclude
that each $\Psi^\dagger$ will contract with some number of 
$A^\dagger$s and that the resulting object will have its single lower 
index contracted with an (upper index) epsilon tensor. 

On the other hand, consider a given upper index on an epsilon tensor. 
This will be contracted either with a $\Psi^\dagger$ or with the lower 
index on an $A^\dagger$, leaving another upper index. In the second 
case, the new upper index will either contract with a $\Psi^\dagger$ or 
another $A^\dagger$, etc..., so it is clear that a given upper index on 
an epsilon tensor will contract with some number of $A^{\dagger}$s and 
then with a $\Psi^\dagger$. 

The previous two paragraphs show that the lower indices on the $Nk$ 
$\Psi^\dagger$s and the upper indices 
on the epsilon tensors (there must be $k$ of them) must pair up 
completely (with arbitrary numbers of intermediate $A^\dagger$s) giving 
$k$ blocks of the form  
\be
\label{block}
\epsilon^{i_1 \cdots i_N} (\Psi^\dagger A^{\dagger}{}^{n_1})_{i_1} 
\cdots 
(\Psi^\dagger A^{\dagger}{}^{n_N})_{i_N} \; .
\ee
Because of the antisymmetry of the $\epsilon$ tensor, this expression 
will vanish unless all of the $n_i$'s are different, so without loss of 
generality, we may take $0 \le n_1 < \cdots < n_N$ in each of the 
blocks.   

We have now accounted for all $\Psi^\dagger$s and epsilon tensors, but 
we may have some additional $A^\dagger$s with indices contracted amongst 
themselves. Such terms will be generated by products of ${\rm 
Tr}(A^{\dagger}), \dots, {\rm Tr}(A^\dagger{}^N)$.\footnote{Note that 
$\det(A^\dagger)$ as well as traces of higher powers of $A^{\dagger}$
may be written in terms of these.} Thus, a given singlet wavefunction 
may also include a set of $A^{\dagger}$s of the form
\be
\label{Ablock}
(\tr{A^\dagger {}^N})^{c_N} \cdots (\tr{A^\dagger})^{c_1} \; .
\ee

To summarize, a general $SU(N)$ singlet energy eigenstate wavefunction 
will be built out of terms with identical numbers of $A^\dagger$s each 
containing $k$ blocks of the form (\ref{block}) plus a block of the 
form (\ref{Ablock}) acting on the Fock space vacuum. Actually, this 
description is rather redundant, and using Fact 3 from the appendix, it 
is straightforward to obtain the following two minimal bases of these singlet 
energy eigenstate wavefunctions. 

The first basis is obtained by restricting to $c_i=0$ in 
(\ref{Ablock}), taking $n_i = i - 1$ in $k-1$ of the blocks 
(\ref{block}) and taking $n_i$ arbitrary (but distinct) in the $k$th 
block to get
\be
\label{basis1}
|\{n_i\}, k \rangle =\epsilon^{i_1 \cdots i_N}
 (\Psi^\dagger A^{\dagger}{}^{n_1})_{i_1} \cdots 
(\Psi^\dagger A^{\dagger}{}^{n_N})_{i_N}
 \left( \epsilon^{i_1 \cdots i_N} (\Psi^\dagger 
A^\dagger 
{}^0)_{i_1}
\cdots (\Psi^\dagger A^\dagger {}^{N-1})_{i_N} \right)^{k-1} 
 |0\rangle \ ; .
\ee
where $0 \le n_1 < \cdots < n_N$. 
From (\ref{ham}), we find that these states have energy levels 
$E(\{n_i\}, k) = \omega({N \over 2}(N+(k-1)(N-1)) + \sum n_i)$.

An alternative basis is obtained by fixing $n_i$ at their minimum 
values $n_i = i - 1$ in each block (\ref{block}) and taking $c_i$ 
arbitrary in (\ref{Ablock}),
\be
\label{basis2}
|\{c_i\}, k\rangle = (\tr{A^\dagger {}^N})^{c_N} \cdots 
(\tr{A^\dagger})^{c_1} 
\left( \epsilon^{i_1 \cdots i_N} (\Psi^\dagger A^\dagger {}^0)_{i_1} 
\cdots (\Psi^\dagger A^\dagger {}^{N-1})_{i_N} \right)^k |0\rangle \; .
\ee
The energy levels for these states are given by $E(\{c_m\}, k) = 
\omega({N \over 2}(N+k(N-1)) + \sum m c_m)$. It is not difficult to
verify that these energy levels and degeneracies precisely match those
of (\ref{basis1}).

Each of these bases realize precisely the known energy levels and 
degeneracies of the model (given in \cite{poly} and references 
therein), so we may conclude that each gives a minimal basis of 
linearly independent energy eigenstate wavefunctions.

We now compare the states (\ref{basis1}) and (\ref{basis2}) to the 
Laughlin-type wavefunctions (\ref{ba1}) and (\ref{ba2}) respectively 
that we constructed in section 2. We see that the states have been 
labeled in an identical fashion, and that states with the same label 
have the same energy level relative to the respective ground states 
(with the identification $\omega = {\kappa \over B}$). More strikingly, the 
similarly labeled states have almost precisely the same form. In particular,
 the formal substitution $\Psi^\dagger_i 
\to 1, A^\dagger_{ij} \to \delta_{ij} a^\dagger_i$ maps a given state 
in (\ref{basis1}) or (\ref{basis2}) precisely to the similarly labeled 
state in (\ref{ba1}) or (\ref{ba2}). 

This comparison makes it obvious that the level $k$ finite matrix
Chern-Simons theory has exactly the same energy levels and the
same number of states at each energy level as the theory of $N$
composite fermions in the lowest Landau level. This
guarantees that the two theories are equivalent, since for each energy
level, we can find an orthogonal basis of wavefunctions in each of the
theories and then define an isomorphism that maps the orthogonal bases
into each other. In this way, we generate an isomorphism between the
Hilbert spaces in the two theories that preserves the inner product
structure and maps the respective Hamiltonians into each other. In
other words, the finite $N$, level $k=2p+1$ matrix Chern-Simons theory is
precisely equivalent to the theory of $N$ (noninteracting) composite 
fermions with $2p$
attached flux lines in the Lowest Landau level, which is believed to 
accurately 
describe the filling fraction $1/k$ state of $N$ electrons and its 
excitations.\footnote{The equivalence has been established with the
inclusion of a harmonic oscillator potential in each model to break
the translation invariance, but the strength of this may be taken to
zero if desired.}

It should be noted that despite the formal similarity between the
bases (\ref{basis1}), (\ref{basis2}) and (\ref{ba1}), (\ref{ba2}), the
similarly labeled states do not precisely map into one another under the
isomorphism. This may be seen by comparing inner products of similarly
labeled states in the two theories.\footnote{An additional subtlety,
pointed out to us by A. Polychronakos, is that because of the well
known quantum mechanical level shift for Chern-Simons theory, the
matrix Chern-Simons model at level $k$ should be identified with the
quantum Hall states at filling fraction $1/(k+1)$, rather than $1/k$.}
However, the equivalence between energy levels and degeneracies
guarantees that an isomorphism exist. Given the explicit wavefunctions we have
derived, it is possible to construct such an isomorphism explicitly.

In any case, the ground state wavefunctions map into one another, so
the Laughlin wavefunction for filling fraction $1/(k+1)$ (see previous 
footnote) is equivalent to the wavefunction
\[
|0_k\rangle = \left( \epsilon^{i_1 \cdots i_N} (\Psi^\dagger A^\dagger 
{}^0)_{i_1} \cdots (\Psi^\dagger A^\dagger {}^{N-1})_{i_N} \right)^k 
|0\rangle
\]
of the matrix Chern-Simons theory. 

It is interesting to note that the equivalence we have described is
related to the work of \cite{bhv,bhkv} on the Calogero model
\cite{calogero}. They
showed that the Calogero model for a certain range of parameters could
be identified with the physics of anyons (particles with fractional 
statistics) in the lowest Landau level (originally conjectured in
\cite{hlm}). On the other hand, it has been shown \cite{op,poly3} 
that the finite
matrix models we have considered are equivalent to the Calogero model
with certain discrete values of the parameters (though not values
appropriate for describing anyons). Thus, the equivalence we have 
demonstrated is related to two
previously known equivalences, though our system does not contain
particles with fractional statistics, but rather composite fermions
with a sort of ``higher order'' integer statistics (higher order zeros
in the wavefunction when particles become coincident).

\section{Discussion}

We have established a precise equivalence between the finite $N$
matrix Chern-Simons theory and the theory of $N$ composite fermions in
the lowest Landau level, believed to accurately describe the filling
fraction $1/k$ fractional Quantum hall state of $N$ electrons. To be
precise, the latter theory is defined to be the theory of $N$
non-interacting fermions in the lowest Landau level with the constraint
that the wavefunctions have a zero of order $\ge k$ whenever two of the
particles become coincident. 

In the large $N$ limit, the composite fermion theory should provide a
good description of the $\nu = 1/k$ quantum Hall fluid of infinite
extent, since the boundary of the quantum Hall droplet goes off to infinity.
On the other hand, as discussed in section 4, we expect that the large
$N$ matrix Chern Simons theory (with the harmonic oscillator potential
turned off) should be equivalent to the original noncommutative
Chern-Simons field theory. Thus, as conjectured by Susskind, the
noncommutative field theory at level $k$ provides a good description
of the of the filling fraction $1/k$ state of electrons in a strong
magnetic field, and in particular is precisely equivalent to the
Laughlin theory (and its present understanding as describing
noninteracting composite fermions in the lowest Landau level).

\section*{Acknowledgements}
 
We would especially like to thank Lenny Susskind for many valuable
discussions out of which this project originated. 
We have also benefitted from discussions with
Michal Fabinger, Matt Kleban, John McGreevy, and Nick Toumbas. M.V.R. would
like to thank the Institute for Theoretical Physics in Santa Barbara
for hospitality during the final stages of this work.  
The work of M.V.R. is supported in part by the Stanford Institute 
for Theoretical Physics and by NSF grant 9870115. The work of Simeon
Hellerman is supported by the D.O.E. under contract DE-AC03-76SF00098 and by 
a D.O.E. OJI grant.    

\appendix

\section{Facts to obtain a minimal basis of wavefunctions.}

In section 5, we derived an exhaustive set of wavefunctions
describing energy eigenstates of the finite matrix Chern-Simons
theory. In this appendix, we derive a number of facts that help to 
determine a minimal basis of these wavefunctions.\\
\\
{\bf Fact 1:}
Defining
\[
A(z) \equiv \prod_{i < j} (z_i - z_j)
\]
we have 
\[
A(z) = \epsilon^{i_1 \cdots i_N} z^0_{i_1} \cdots z^{N-1}_{i_N}
\; .
\]
\\
Proof: From the definition, it is clear that $A$ is the lowest order
polynomial that is completely antisymmetric in its arguments, since any such polynomial must have a factor $(z_i-z_j)$ for each $i \ne j$. The
second expression is also completely antisymmetric and has the same
order, so it must equal $A$ up to a numerical factor which is easily
checked to be 1.\\
\\
{\bf Fact 2:}
Any polynomial 
\be
\label{form1}
f(z) = \epsilon^{i_1 \cdots i_N} z_{i_1}^{n_1} \cdots 
z_{i_N}^{n_N} 
\ee
may be written as a sum of terms of the form
\be
\label{form2}
g(z) = S_N^{c_N} \cdots S_1^{c_1} A(z) 
\ee
where $S_k = \sum_i z_i^k$. Conversely, any polynomial of the form
(\ref{form2}) may be written as a sum of terms of the form 
(\ref{form1}).\\ 
\\
Proof: The polynomial $f(z)$ is completely antisymmetric in its
arguments, thus it vanishes when any two of its arguments are set
equal. This implies that $A$ divides $f$, so $f = SA$ for some
polynomial $S$. Since $f$ and $A$ are both totally antisymmetric, $S$
must be totally symmetric. It is not difficult to show that any 
symmetric 
polynomial $S(z_1,\dots,z_N)$ may be written uniquely as a sum of
terms of the form
\[
S_N^{c_N} \cdots S_1^{c_1} 
\]
where $S_k = \sum_i z_i^k$. To prove the converse, we simply note
that polynomials of the form (\ref{form2}) are completely
antisymmetric and are thus allowed coherent state wavefunctions for n
fermions in a harmonic oscillator potential. On the other hand, as
pointed out in section 2, polynomials of the form (\ref{form1})
represent the coherent state wavefunctions for an orthogonal basis of
such states.\\
\\ 
{\bf Fact 3:} 
Let $\Psi^\dagger$ and $A^\dagger$ be an $N$-dimensional vector and an
$N \times N$  matrix of commuting variables. Then any expression of the
form 
\be
\label{form3}
F(\Psi^\dagger, A^\dagger) = \epsilon^{i_1 \cdots i_N} 
(\Psi^\dagger A^{\dagger}{}^{n_1})_{i_1} \cdots 
(\Psi^\dagger A^{\dagger}{}^{n_N})_{i_N}
\ee
may be written uniquely as a sum of terms of the form 
\be
G(\Psi^\dagger,A^\dagger) = (\tr{A^\dagger {}^N})^{c_N} \cdots
 (\tr{A^\dagger})^{c_1} 
\epsilon^{i_1 \cdots i_N} (\Psi^\dagger A^\dagger {}^0)_{i_1} 
\cdots (\Psi^\dagger A^\dagger {}^{N-1})_{i_N} 
\label{form4}
\ee
Conversely, any expression of the form (\ref{form4}) may be written
uniquely as a sum of terms of the form (\ref{form3}). \\
\\
Proof: First note that there is a one-to-one correspondence between 
expressions
of the form (\ref{form3}) and polynomials of the form (\ref{form1})
obtained by taking the same $n_i$, and a one-to-one correspondence
between expressions of the form (\ref{form4}) and polynomials of the
form (\ref{form2}) obtained by taking the same $c_i$. By Fact 2, given 
any
$F$, the corresponding polynomial $f$ may be written uniquely as a sum
of terms of the form (\ref{form2})
which we denote $\sum g_i$. We now prove that 
\be
\label{f3}
F(\Psi^\dagger, A^\dagger) = \sum G_i(\Psi^\dagger, A^\dagger)
\ee
where $G_i$ is the expression corresponding to $g_i$. First, note that
for $A^\dagger_{ij} = \delta_{ij} z_i$ we have 
\be
\label{step2}
F(\Psi^\dagger, A^\dagger) = f(z_i) \prod \Psi^\dagger_j = \sum 
g_i(z_i) \prod
\Psi^\dagger_j = \sum G_i(\Psi^\dagger, A^\dagger) \; .
\ee
Next, let $A^\dagger$ be an arbitrary hermitian matrix of complex
numbers. Then $A^\dagger = U D U^\dagger$ for some unitary matrix $U$
and real diagonal matrix $D$. It is straightforward to check using 
(\ref{step2}),
\[
F(\Psi^\dagger, A^\dagger) = F(\Psi^\dagger U, D) = \sum
G_i(\Psi^\dagger U, D) = \sum G_i(\Psi^\dagger, A^\dagger) \; .
\]
Now, we have shown that (\ref{f3}) holds for arbitrary $\Psi^\dagger$
and any hermitian matrix $A^\dagger$. But both sides of (\ref{f3})
are simply finite polynomials in the components of $\Psi^\dagger$ and
$A^\dagger$ (and not their Hermitian or complex conjugates). Therefore
the equivalence clearly cannot depend on $A^\dagger$ being Hermitian
so it must hold for any numerical matrix $A^\dagger$ and vector
$\Psi^\dagger$. This implies that the two polynomials are equivalent
for all numerical values of their arguments, so we may conclude that
(\ref{f3}) is true as an identity between polynomials of commuting
variables. The fact that the decomposition (\ref{f3}) is unique
follows immediately from the uniqueness in fact 2. Thus, any $F$ can
be written uniquely as a sum of $G$s, and by an identical argument, we
can show that any $G$ can be written uniquely as a sum
of $F$'s.

\end{document}